\documentclass[aps,pre]{revtex4}

\usepackage{graphicx}  % got figures? uncomment this
\usepackage{epsfig}
\usepackage{dcolumn}% Align table columns on decimal point
\usepackage{bm}% bold math
\usepackage{amsmath}% if you are using this package, it must be loaded before amsthm.sty
\usepackage{amsthm}
\usepackage{amssymb}

\begin{document}

\newcommand{\hilight}[1]{\colorbox{yellow}{#1}}
\newcommand{\rbold}{\ensuremath{\textbf{r}}}
\newcommand{\Rbold}{\ensuremath{\textbf{R}}}
\newcommand{\ombold}[1]{\ensuremath{\mathbf{\omega}_{#1}}}
\newcommand{\gr}[1]{\ensuremath{g(#1)}}
\newcommand{\Gr}[1]{\ensuremath{G(#1)}}
\newcommand{\gtwor}{\ensuremath{g_{2}(\rbold_{1},\rbold_{2})}}
\newcommand{\gthreer}{\ensuremath{g_{3}(\rbold_{1},\rbold_{2},\rbold_{3})}}
\newcommand{\dirac}[1]{\ensuremath{\delta(#1)}}
\newcommand{\kronecker}[1]{\ensuremath{\delta_{#1}}}
\newcommand{\hr}[1]{\ensuremath{h(#1)}}
\newcommand{\rmin}{\ensuremath{r_{min}}}
\newcommand{\rmax}{\ensuremath{r_{max}}}
\newcommand{\gabr}[2]{\ensuremath{g_{#1}(#2)}}
\newcommand{\Gabr}[2]{\ensuremath{G_{#1}(#2)}}
\newcommand{\habr}[2]{\ensuremath{h_{#1}(#2)}}
\newcommand{\coordno}[2]{\ensuremath{N(#1,#2)}}
\newcommand{\coordabno}[3]{\ensuremath{N_{#1}(#2,#3)}}
\newcommand{\ca}[1]{\ensuremath{c_{#1}}}
\newcommand{\caa}[1]{\ensuremath{c_{#1}^2}}
\newcommand{\bbar}[1]{\ensuremath{\left\langle b_{#1}\right\rangle}}
\newcommand{\bbartwo}[1]{\ensuremath{\left\langle b_{#1}^{2}\right\rangle}}
\newcommand{\dsdo}[2]{\ensuremath{\left(\dfrac{d\sigma}{d\Omega}\right)^{#1}_{#2}(\lambda,2\theta)}}
\newcommand{\dsdosimp}[2]{\ensuremath{\left(\dfrac{d\sigma}{d\Omega}\right)^{#1}_{#2}}}
\newcommand{\fofq}[3]{\ensuremath{F_{#1}^{#2}(#3)}}
\newcommand{\qbold}{\ensuremath{\textbf{Q}}}
\newcommand{\sq}[1]{\ensuremath{S(#1)}}
\newcommand{\hq}[1]{\ensuremath{H(#1)}}
\newcommand{\sabq}[2]{\ensuremath{S_{#1}(#2)}}
\newcommand{\habq}[2]{\ensuremath{H_{#1}(#2)}}
\newcommand{\fxq}[1]{\ensuremath{f_{#1}(Q)}}
\newcommand{\fxqzero}[1]{\ensuremath{f_{#1}(0)}}
\newcommand{\fxqtwo}[1]{\ensuremath{f_{#1}^2(Q)}}
\newcommand{\fxqtwozero}[1]{\ensuremath{f_{#1}^2(0)}}
\newcommand{\bnbar}[1]{\ensuremath{\left\langle b_{#1}\right\rangle}}
\newcommand{\bnbartwo}[1]{\ensuremath{\left\langle b_{#1}^2\right\rangle}}
\newcommand{\dlmn}[2]{\ensuremath{D^{l_{#1}}_{m_{#1}#2}(\ombold{#1})}}
\newcommand{\dlmnmod}[3]{\ensuremath{D^{l_{#1}}_{m_{#1}#2}(\ombold{#3})}}
\newcommand{\TOF}{\ensuremath{\text{TOF}}}

\newcommand{\Phiavt}{\ensuremath{\langle \Phi(\lambda,t)\rangle_{t}}}
\newcommand{\DeltaOmega}{\ensuremath{\Delta\Omega(2\theta)}}

\title{Water and ice structure in the range 220 - 365K from radiation total scattering experiments}

\author{Alan~K.~Soper}

\affiliation{ISIS Facility, STFC Rutherford Appleton Laboratory, Harwell Oxford, Didcot, OX11 0QX, UK}

%\institute{ISIS Facility, STFC Rutherford Appleton Laboratory, Harwell Oxford, Didcot, OX11 0QX, UK}
%% When only one author is present, please do not use the command \from{} near the author name.

\date{\today}

\begin{abstract}
The past decade or so has witnessed a large number of articles about water structure. The most incisive experiments involve radiation with a wavelength compatible with the observed inter-molecular separations found in water, of order $\sim 3$\AA, in other words mostly $<$1eV neutrons and $>$10keV x-rays. Because x-rays are scattered by electrons while neutrons are scattered by nuclei, the two probes give complementary information, which, when combined with the pronounced isotopic contrast available between deuterons and protons, enables experiments to be devised that allow the three site-site radial distribution functions for water, namely O-O, O-H and H-H, to be determined uniquely. In practice systematic effects in both neutron and x-ray experiments prevent this ideal being attained, so recourse is made to computer simulation to extract these distribution functions from the data. Here a flavour of Monte Carlo simulation called Empirical Potential Structure Refinement (EPSR for short) is used to devise an empirical intermolecular potential which attempts to drive the simulated radial distribution functions as close as possible to the data. New x-ray and neutron scattering data on water in the temperature range 280 - 365K are presented for the first time, alongside a new analysis of some much older neutron data on ice 1h at 220K. This temperature analysis, above and below the water freezing point of water, reveals some non-intuitive water properties in the liquid and solid states.
\end{abstract}

\maketitle

\section{Introduction}

There continues to be extended activity trying to understand the structure of water. This activity arises naturally from the fact that water is essential to the existence of life on Earth, it is a major component in human industry and welfare, and its properties have a profound impact on the atmosphere and climate that surrounds us. The demand for structural data stems from the need to try to understand water properties at the molecular level: these are determined by the forces of interaction between and within water molecules, but the search for a reliable force field that quantitatively reproduces all known water properties seems to remain as illusory as ever. Which is not to say there has not been a huge amount of effort and progress in this field.

Broadly speaking, water force fields can be categorised into three main kinds. There are the qualitative force fields: these are often relatively simple in mathematical form and are designed to study a water-like fluid over a range of thermodynamic state points. Examples of such force fields include, but are not by any means limited to, the ST2 potential of Rahman and Stillinger, \cite{rahman1971molecular}, the SPC potential \cite{berendsen1981interaction}, the earlier versions of the TIP series of potentials \cite{jorgensen1981quantum} or the mW potential \cite{molinero2008water}. The second set, also often of simpler mathematical form but not necessarily so, aim to give a better quantitative description of water structure and thermodynamic properties, at the expense of some ease of use in terms of the computational time needed to run them. Examples of these potentials include the TIP4P2005 potentials \cite{abascal2005general}, and those potentials where quantum mechanical effects are calculated explicitly as part of the simulation, rather than as part of the empirical potential, \cite{TTM3-F2008, fanourgakis2009fast}. Finally there are the \textit{ab initio} methods where Molecular Dynamics is combined with Density Functional Theory to attempt to capture the electronic interaction between water molecules. Naturally these have to be run on much smaller systems than is possible with the simpler potentials, but the results appear to be dependent on the choice of density functional and are often found to be too structured and to have too slow dynamics compared to experiment and the empirical potentials.\cite{khueneparrinello2009} Obviously the list of potentials and methods here is far from exhaustive and is not intended to be so: the point is there is very diverse range of potentials and methods available because none of them is completely satisfactory.

Extracting reliable experimental information on the structure of water, via the site-site radial distribution functions, \gabr{OO}{r}, \gabr{OH}{r} and \gabr{HH}{r}, proves to be equally problematic, as highlighted in a recent review, \cite{soper2013radial}. As documented in that work there have been numerous attempts at the x-ray scattering experiment on water with varying degrees of consensus on the different results, and the same is certainly true of the neutron scattering approach. However it does seem that independent approaches to this problem are converging to a common picture: compare, for example, the results from \cite{hura2003water} with those of \cite{soper2000}, or the results of \cite{skinner2013benchmark} with those of \cite{soper2013radial}. Whilst these latest results are certainly not error free, it seems that, by comparing many different experiments, a measure of the true uncertainties in the extracted radial distribution functions is obtained.

This of course all applies to ambient water. The situation for non-ambient water is far less complete and even today there are still only a handful of scattering experiments on liquid water which study the effect of changing temperature and pressure. See for example \cite{narten1971:1}, \cite{okhulkov1994x}, \cite{yamanaka1994structure}, \cite{soper2000}, \cite{soper2000structures}, \cite{strassle2006structure}, and \cite{katayama2010structure}. Therefore in an attempt to correct this imbalance between ambient and non-ambient water, I show here a new analysis of some previous neutron scattering data on water in the temperature range 10 - 365K \cite{soper2011water}, now combined with new x-ray data over the same temperature range. At the same time the opportunity is taken to re-analyse some earlier neutron scattering data on ice Ih at 220K \cite{soper2000}. The comparison of the liquid at different temperatures with ice Ih structure produces an intriguing story.

\section{\label{experiment}Experimental}

\subsection{\label{xray}X-ray scattering experiment}
X-ray scattering data for water at 295K, ambient pressure, were recorded on a PANalytical x-ray diffractometer, using the white x-ray beam from a silver anode (K$_{\alpha}$ wavelength = 0.5609\AA) using a 2.5mm silica glass thin walled capillary in transmission geometry. A focussing mirror was used to eliminate as much off-K$_{\alpha}$ radiation as possible. The sample temperature was controlled by means of a Cryostream heater and cooler to control the temperature within 0.1K. Temperature uniformity across the length of the capillary was not monitored but was expected to be not worse than $\sim$2K. The scattering data were corrected for background, empty capillary scattering, attenuation, multiple scattering and Compton scattering, and put on an absolute scale of electron units using the Krogh-Moe method \cite{kroghmoe}. In addition, using diffraction data from silicon crystalline powder as a calibrant, a correction was developed to remove the off-energy scattering that arose from the significant bremsstrahlung radiation in the incident x-ray spectrum. Finally the single atom scattering was subtracted from the diffraction data, which were then normalised to the same single atom scattering to give an x-ray interference differential scattering cross section (Fig. 2(a)), defined by:

\begin{equation}
F_{x}(Q)=\frac{\sum_{\alpha\beta\ge\alpha}\left( 2-\kronecker{\alpha\beta}\right) c_{\alpha}c_{\beta}\fxq{\alpha}\fxq{\beta} \habq{\alpha\beta}{Q}}{\sum_{\alpha}c_{\alpha}\fxq{\alpha}^{2}}
\label{xrddata}
\end{equation}
where $c_{\alpha}$ is the atomic fraction and $\fxq{\alpha}$ is the atomic form factor for component $\alpha$, and the partial structure factor, $\habq{\alpha\beta}{Q}$ between atom types $\alpha$ and $\beta$ is defined as the Fourier transform of the corresponding site-site radial distribution function, $\gabr{\alpha\beta}{r}$:
\begin{equation}
\habq{\alpha\beta}{Q}=4 \pi \rho \int r^{2}(\gabr{\alpha\beta}{r}-1)\frac{\sin Qr}{Qr}dr
\label{psf}
\end{equation}
with $\rho$ the atomic number density and $Q$ the wave vector change in the scattering experiment.

\subsection{\label{neutron}Neutron scattering experiment}

Neutron scattering gives fundamentally the same information as the x-ray experiment, except that atomic form factors are replaced by numbers - neutron scattering lengths - one for each isotope. This means that heavy water has a completely different scattering profile compared to normal light water. This can be exploited by measuring heavy and light water, and mixtures thereof, to give direct information on the H-H and O-H correlations in the liquid. In the present instance, as well as the pure liquids, measurements were made on mixtures of 50 mole\%\ and 64 mole\%\ H$_2$O in D$_2$O, the latter sample being called ``null'' in the figures because the net coherent scattering length of hydrogen in this sample is close to zero. Scattering data were corrected for background scattering, container (made from TiZr alloy) scattering, attenuation, multiple scattering, inelastic scattering (using the methods described in \cite{soper2013radial}), and put on an absolute scale by comparison with the scattering from a known volume of vanadium, which has an almost incoherent scattering cross section for neutrons. The samples were mounted on a temperature controlled sample changer and heated or cooled by means of a circulating water bath. 

The resulting interference differential scattering cross section (Fig. 2(a)) for neutrons is (assuming all the corrections have been performed perfectly):
\begin{equation}
F_{n}(Q)=\sum_{\alpha\beta\ge\alpha}\left( 2-\kronecker{\alpha\beta}\right) c_{\alpha}c_{\beta}\bbar{\alpha}\bbar{\beta} \habq{\alpha\beta}{Q}
\label{neutdata}
\end{equation}
where the angular brackets represent averages over the spin and isotope state of the respective nuclei. The neutron scattering data in this work were recorded as part of the commissioning experiments of the new NIMROD diffractometer at ISIS which is designed for looking at intermediate range structure in liquids, complex fluids, and glasses \cite{NIMROD}. Comparison of these datasets with those measured previously on the SANDALS diffractometer at ISIS, \cite{soper2000,soperjpcm2007} shows excellent agreement in general, though problems with inelastic scattering from light hydrogen appear to be more pronounced on NIMROD due to the smaller ratio of scattered flight path to incident flight path compared to SANDALS.

For the ice data a different sampling protocol is required. The point is that if a container of water is simply frozen, the crystalline material will contain a very high degree of preferred orientation, making obtaining a reliable powder average from such material difficult. (The flat plate containers used in these experiments cannot be easily rotated to reduce this effect.) One way to avoid this is to freeze the sample, then powder it manually before placing in the container. This eliminates the preferred orientation, but also means we have no absolute scale for the scattering data, which become dependent on an unknown `packing fraction''. To circumvent both problems in the present case a series of samples of  H$_2$O, D$_2$O and ``null'' ice were measured by repeated freezing and thawing cycles: this helped to obtain a reasonable powder average but may not have removed all the preferred orientation. 

\subsection{\label{EPSR}Empirical Potential Structure Refinement}

Empirical potential structure refinement (EPSR) is a Monte Carlo computer simulation method to simulate distributions of atoms and molecules which are consistent with a set of radiation scattering data. The method has been described extensively in previous publications \cite{soper2005,soper2013radial} so will not be elaborated here, other than to say the method uses an existing molecular interaction potential, the reference potential, to create an initial distribution of molecules in the box. Then a perturbation to this potential is derived from the difference between simulated and measured structure factors, to drive the simulated distributions as close as possible to the measured data. The reference potential used in the present work was a Lennard-Jones plus Coulomb plus exponential repulsive potential of the form:
\begin{equation}
U_{\alpha\beta}^{(ref)}(r)=4\epsilon_{\alpha\beta}\left[\left(\dfrac{\sigma_{\alpha\beta}}{r}\right)^{12}-\left(\dfrac{\sigma_{\alpha\beta}}{r}\right)^{6}\right]+\dfrac{q_{\alpha}q_{\beta}}{4\pi\epsilon_{0}r}+C_{\alpha\beta}\exp\left(\frac{1}{\gamma}\left(R_{\alpha\beta}-r\right)\right)
\label{eq:refpot}
\end{equation}
where $\epsilon_{\alpha\beta}=\sqrt{\epsilon_{\alpha}\epsilon_{\beta}}$, $\sigma_{\alpha\beta}=0.5\left(\sigma_{\alpha}+\sigma_{\beta}\right)$ are the well depth and hard-core distance for atom pair $\alpha\beta$ respectively, $q_{\alpha}$ is the charge on atom type $\alpha$, $R_{\alpha\beta}$ is an additional pre-assigned minimum distance for atom pair $\alpha\beta$, to compensate for any atomic overlaps caused by the empirical potential, and $\gamma$ is a width parameter which controls the hardness of the repulsive potential.The coefficient $C_{\alpha\beta}$ adjusts automatically depending on whether atom pairs of type $\alpha,\beta$ are found for $r<R_{\alpha\beta}$. In all the simulations presented here, $\gamma$ was set to 0.3\AA

For the simulations of crystalline ice, an additional harmonic potential was introduced. It was applied to the centre of mass of each water molecule. The force constant used in this ``tethering'' potential was used to control how far away water molecules could drift from their ideal lattice sites, and so stop the the lattice from melting. Increasing this force constant produced a more sharply defined lattice with smaller Debye-Waller factors and weaker diffuse scattering. Apart from this additional tethering potential, the water molecules were free to rotate and move as in the liquid state. 

The parameters used in the reference potential in these simulations are given in Table \ref{tab:refpotpars}. Note that a smaller value of $\sigma_{O}$ is used for ice than for water because the tethering potential already prevents molecules approaching one another too closely, so the use of the Lennard-Jones potential to prevent atomic overlap is redundant. The minimum distances, used in both sets of simulations, are shown in Table \ref{tab:minimumdistances}. These minimum distances are introduced to help improve the overall fit to the scattering data, and to compensate for the action of the empirical potential at very short distances, but have no other significance.

\begin{table}
 \caption{Reference potential parameters used in the EPSR simulations of water and ice}
\label{tab:refpotpars}
  \begin{tabular}{lccc}
    \hline
    Atom & $\epsilon$ & $\sigma$ & q \\
    & [kJ/mol] & [\AA] & [e] \\
    \hline
 Water \\
    \hline
        O &  0.3   & 3.2  & -1.0    \\
        H &   0.0  & 0.0  & +0.5    \\
 	\hline
 	Ice \\
    \hline
        O &  0.3   & 2.2  & -1.0    \\
        H &   0.0  & 0.0  & +0.5    \\
    \hline
  \end{tabular}
\end{table}

\begin{table}
 \caption{Minimum atom-atom distances used in the EPSR simulations of water and ice, equation \ref{eq:refpot}.}
\label{tab:minimumdistances}
  \begin{tabular}{ccc}
    \hline
    Atom pair & $R_{\alpha\beta}$ \\
    &  [\AA] \\
    \hline
 Water \\
    \hline
        O-O & 2.39 \\
        O-H & 0.00 \\
        H-H & 0.00 \\
    \hline
 Ice \\
    \hline
        O-O & 0.00 \\
        O-H & 0.00 \\
        H-H & 1.70 \\
    \hline
  \end{tabular}
\end{table}

The water simulations were performed in a cubic box. Table \ref{tab:simboxes} lists the box dimensions and atomic number densities that were implemented at the six temperatures. For the ice simulations, a hexagonal unit cell was built, based on the lattice parameters given in \cite{rottger1994lattice,rottger2012lattice}, with $a=b=4.5151$\AA\ and $c=7.3509$\AA, giving an atomic number density of 0.092465 atoms/\AA$^{3}$. The simulation box was built from 9 unit cells along each of the $a$ and $b$ axes and 6 unit cells along the $c$ axis, making a total of 1944 water molecules. These values gave generally good agreement with the measured Bragg peak positions in the scattering data. The resolution of the Bragg peaks as a function of $Q$ was determined from a separate measurement on sintered MgO powder. The Bragg peak calculation was performed as described in \cite{soper2013empirical}.

\begin{table}
 \caption{Simulation box sizes and atomic number densities used in the EPSR simulation of liquid water from 280K to 365K. Each simulation used 1000 water molecules.}
\label{tab:simboxes}
  \begin{tabular}{ccc}
    \hline
    Temperature & Box size & Atomic no. density \\
     K &  [\AA] & [atoms/\AA$^{3}$] \\
    \hline
    280 & 31.0413 & 0.1003 \\
    288 & 31.0413 & 0.1003 \\
    295 & 31.0723 & 0.1000 \\
    313 & 31.1139 & 0.0996 \\
    343 & 31.2717 & 0.0981 \\
    365 & 31.4110 & 0.0968 \\
    \hline
  \end{tabular}
\end{table}

For the liquid data the empirical potential was represented as a linear combination of Poisson functions, as described in \cite{soper2005partial}. However, these functions produce a strongly decaying and broadened function for large $r$, by design, which is not entirely suitable for crystal structure refinement where the decay of correlations with increasing $r$ is much weaker. Therefore for the ice structure refinement the empirical potential was represented by a series of Gaussian functions of constant width $\sigma$:
\begin{equation}
p_{n}(r,\sigma)=\frac{1}{4\sqrt{2}\pi^{\frac{3}{2}}\rho rd_{n}}\left[\exp{-\frac{\left(r-d_{n}\right))^{2}}{2\sigma^{2}}}-\exp{-\frac{\left(r+d_{n}\right)^{2}}{2\sigma^{2}}}\right]
\label{eq:pofrgauss}
\end{equation}
 where $d_{n}=n\Delta r$ and $\Delta r$ is the required spacing between points in $r$-space. This form has an exact 3-dimensional Fourier transform to $Q$ space, namely $P_{n}(Q,\sigma)=\frac{\sin Qd_{n}}{Qd_{n}}\exp{-\frac{Q^{2}}{2\sigma^{2}}}$.
 
 The quality of the fit to the data is controlled by the resolution of the empirical potential (EP), which in turn is determined by the spacing between terms, $\Delta r$, and the width $\sigma$ in equation \ref{eq:pofrgauss}, as well as the overall amplitude
\begin{equation}
E_{req}=\sum_{\alpha\beta\ge\alpha} c_{\alpha}c_{\beta}\sum_{r}\vert U_{\alpha\beta}^{(ep)}(r)\vert
\label{eq:ereq}
\end{equation}
The larger the value of $E_{req}$ the more structure in the EP, which may improve the quality of the fit, but which might also introduce artefacts from the data. Hence a balance has to be struck between quality of fit and the presence of artefacts (which may actually make the fit worse). As a general rule $E_{req}$ is increased in steps until there is no further improvement in the quality of fit. For the water simulations shown here, $E_{req}$ was set to 20kJ/mole, while for the ice simulations it was set to 50kJ/mole.
 
\section{Results}

\subsection{\label{results}Fits to the data}

\begin{figure}
\includegraphics{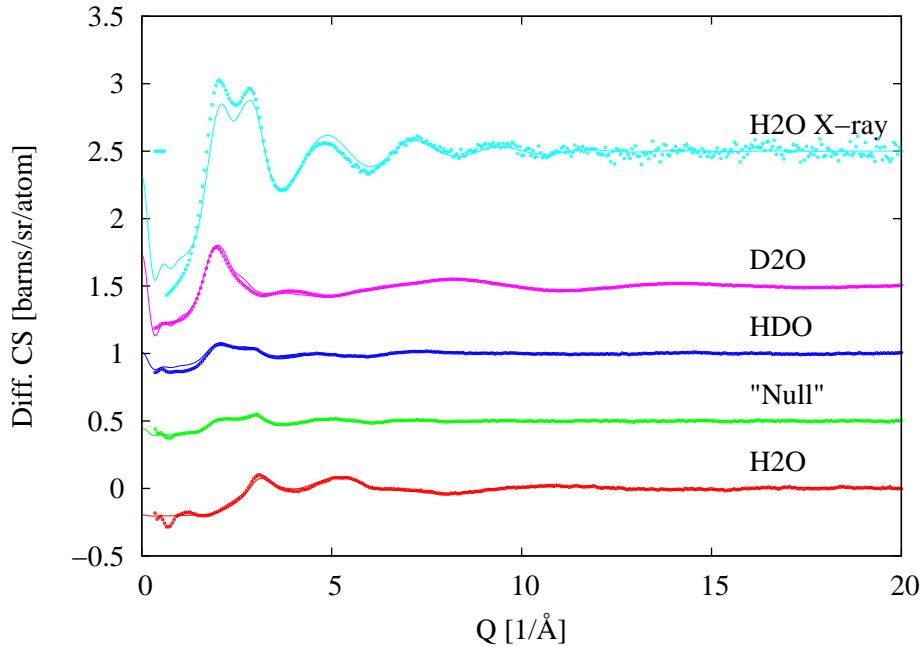} 
\caption{\label{fig:h2o22csqfits}Empirical potential structure refinement (EPSR) fits to neutron scattering data from four mixtures of heavy and light water, together with the EPSR fit to the corresponding x-ray data from light water. The temperature is 295K. Equivalent fits were obtained at all the other temperatures used in this study.}
\end{figure}

\begin{figure}
\includegraphics{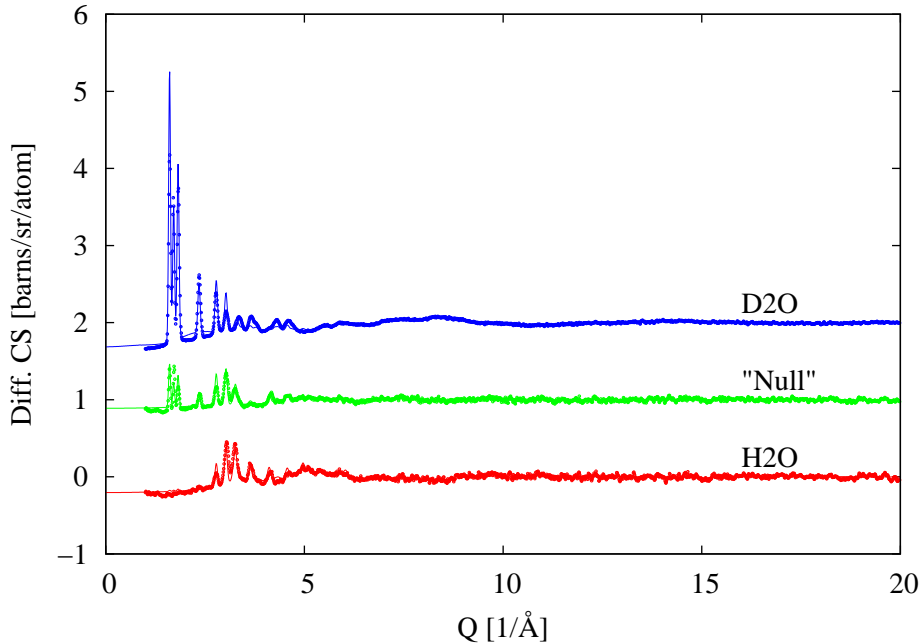} 
\caption{\label{fig:ice220ksqfits}Empirical potential structure refinement (EPSR) fits to neutron scattering data from three mixtures of heavy and light ice. The temperature is 220K.}
\end{figure}

Figure \ref{fig:h2o22csqfits} shows a typical example of fits to the neutron and x-ray liquid water data, while Fig. \ref{fig:ice220ksqfits} shows the corresponding fits to the neutron data for ice at 220K. In both cases it is noticeable that an exact fit to the scattering data is never achieved: there are always residual discrepancies that cannot be accounted for. There are three primary causes for these discrepancies. 

Firstly, of course, the data are themselves not perfect. All the neutron and x-ray scattering data are subject to a series of corrections for attenuation, multiple scattering, background scattering, container scattering, and inelasticity corrections, which arise from nuclear recoil when struck by a neutron or electron recoil when struck by an x-ray. Most of these corrections are straightforward to estimate quite accurately, but the inelasticity corrections, particularly when light atoms like hydrogen are involved, and remain a significant challenge. Errors in calculating the inelasticity corrections give rise to a systematic error in the data which is difficult to estimate accurately. These errors affect all total scattering experiments on liquids, glasses and crystalline materials alike. 

Secondly, the EPSR method is not perfect. No matter how carefully measured, data of these kinds are always subject to Fourier transform errors, which arise from the finite counting statistics and finite $Q$-range of the experiment. Hence the empirical potential in EPSR has to strike a balance between fitting the data as closely as possible, while not fitting artefacts of the Fourier transform and systematic uncertainties. As a result a perfect fit to multiple datasets is always going to be unlikely. Note however that in the present cases the method does appear to have captured the main features of the data, both qualitatively and quantitatively. For the ice data the heights of the Bragg peaks are not always reproduced quantitatively, most likely due to residual preferred orientation in the samples. In addition there appears to be too much diffuse scattering in the model compared to the data in the region $Q=$2\AA$^{-1}$, although it is not clear if this is a genuine feature of the model, or a consequence of the way the fit is estimated from the simulation when Bragg peaks are present \cite{soper2013empirical}.

Finally, it has to be borne in mind that strictly, heavy and light water are not identical structurally \cite{hart2006isotope} so that any attempt to fit the structure data from both liquids and mixtures thereof with a single structural model is ultimately going to end in failure. In practice, of course, the differences between the two liquids are likely to be smaller than the other systematic effects listed above so that observing this difference is a challenge. Compare for example \cite{soper2008quantum} with \cite{zeidler2012isotope}. 

\begin{figure}
\includegraphics{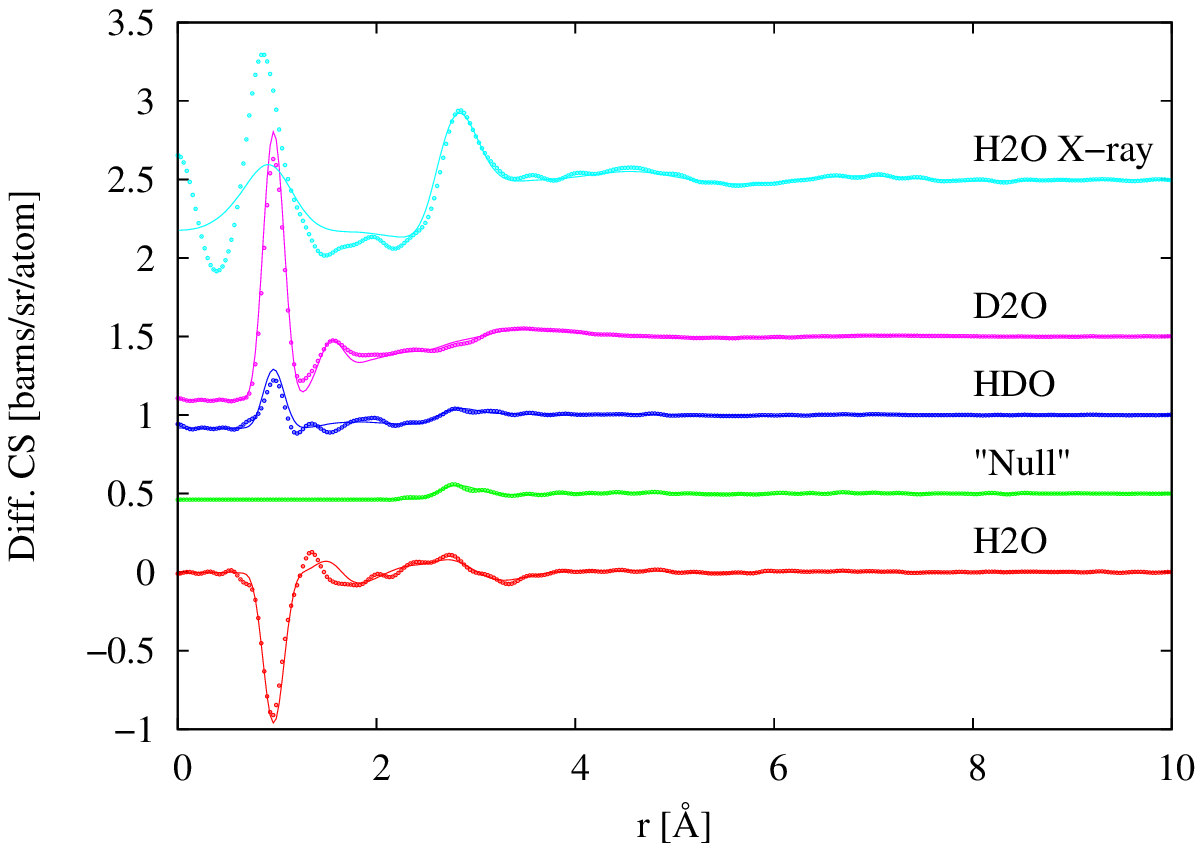} 
\caption{\label{fig:h2o22cgrfits}Real space representation of EPSR fits to neutron and x-ray scattering data from Fig. \ref{fig:h2o22csqfits}. The temperature is 295K. In each case both the data and fit were Fourier transformed over the same $Q$ range.}
\end{figure}

\begin{figure}
\includegraphics{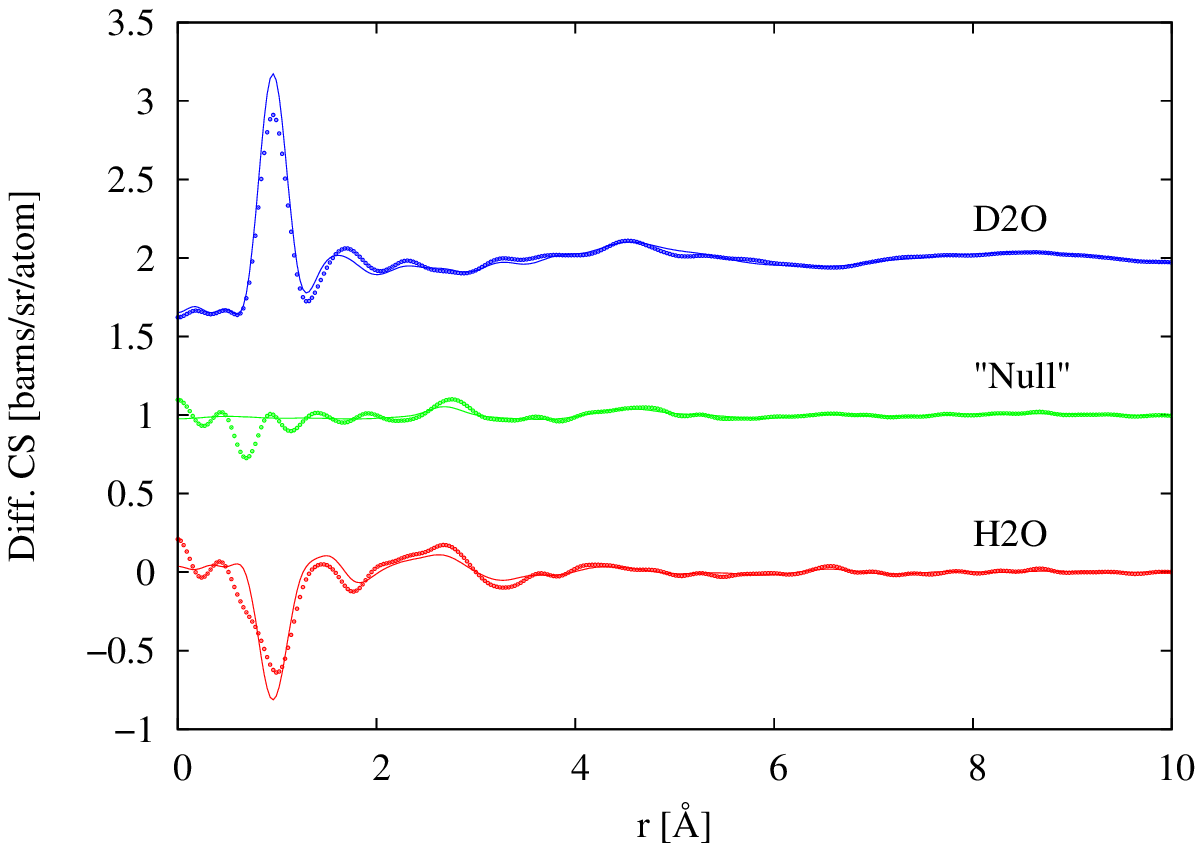} 
\caption{\label{fig:ice220kgrfits}Real space representation of EPSR  fits to the neutron scattering data from three mixtures of heavy and light ice shown in Fig. \ref{fig:ice220ksqfits}. The temperature is 220K.}
\end{figure}

Figures \ref{fig:h2o22cgrfits} and \ref{fig:ice220kgrfits} show the same fits in real space, after Fourier transforming both $Q$-space data and fit over the same range of $Q$ values. Similar comments apply to these fits as for the $Q$-fits, i.e. the fits capture the bulk of the data, but miss-fit slightly in some areas. One notable feature in these $r$-space fits is that in the region of the O-O first peak in $g_{OO}(r)$ near $\sim$2.8\AA, the simulation appears to underestimate the intensity of the scattering data for H$_{2}$O and ``Null'' water, but overestimate it for D$_{2}$O, particularly in the case of ice Ih. However, as can be readily appreciated in these $r$-space representations, it is hard to distinguish between genuine structure and Fourier transform artefact, which is why EPSR only fits to the $Q$-space data.

\begin{figure}
\includegraphics{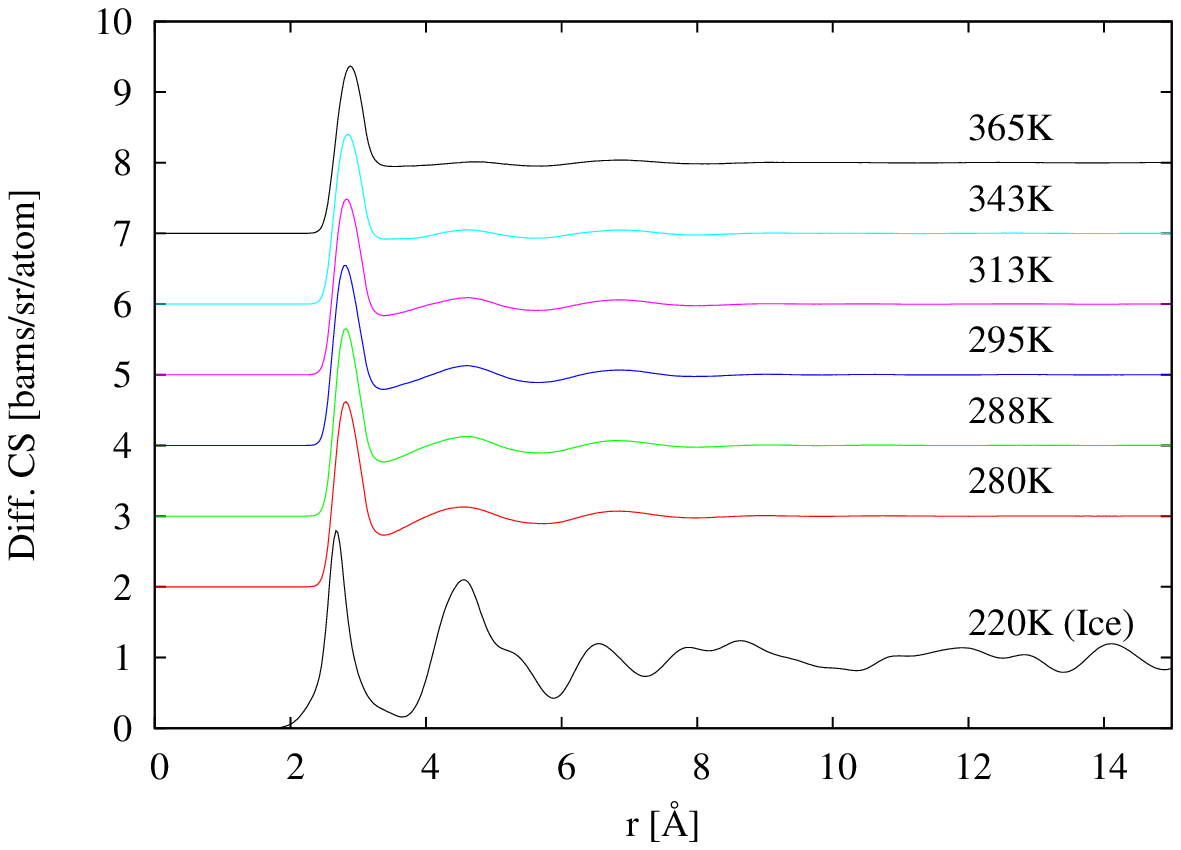} 
\caption{\label{fig:oogrs}OO radial distribution functions for ice at 220K and water in the temperature range 280-365K, as derived from the EPSR simulations shown in Figs. \ref{fig:h2o22csqfits} and \ref{fig:ice220ksqfits}.}
\end{figure}

\begin{figure}
\includegraphics{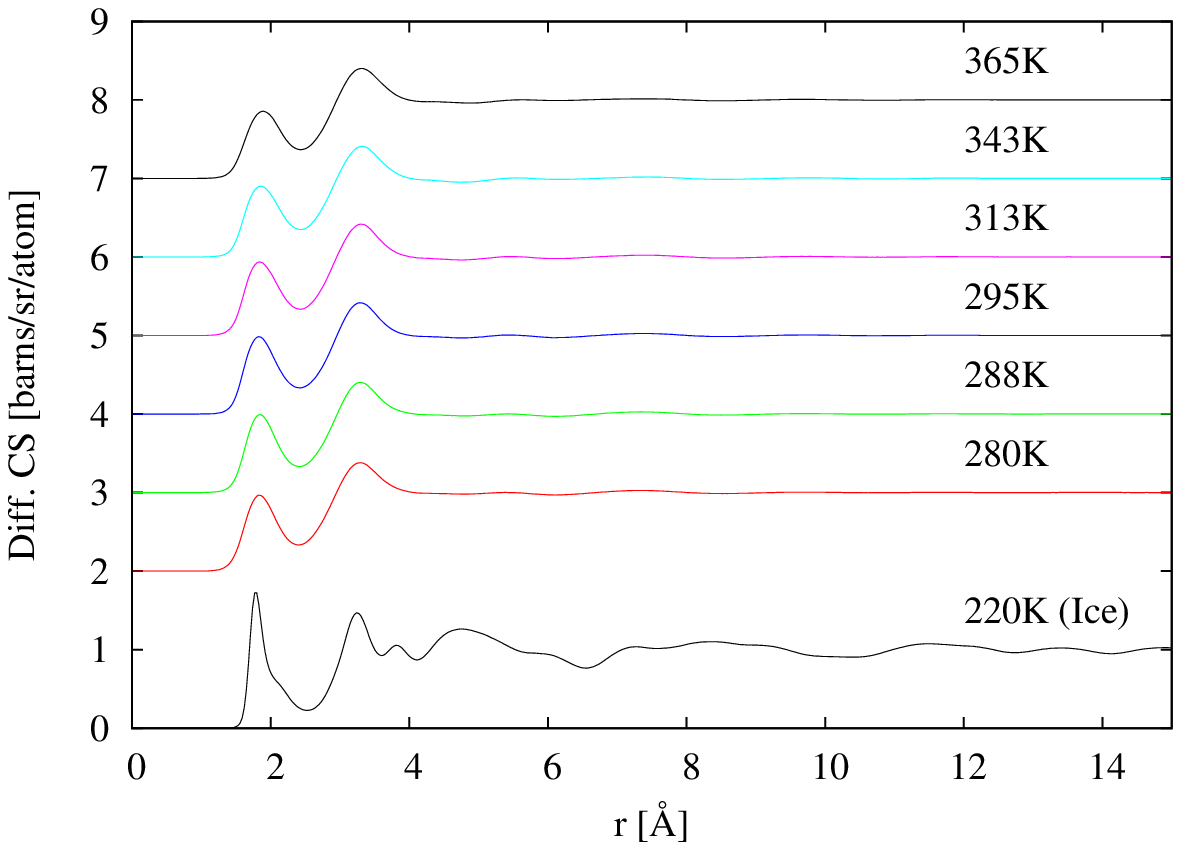} 
\caption{\label{fig:ohgrs}OH radial distribution functions for ice at 220K and water in the temperature range 280-365K, as derived from the EPSR simulations shown in Figs. \ref{fig:h2o22csqfits} and \ref{fig:ice220ksqfits}.}
\end{figure}

\begin{figure}
\includegraphics{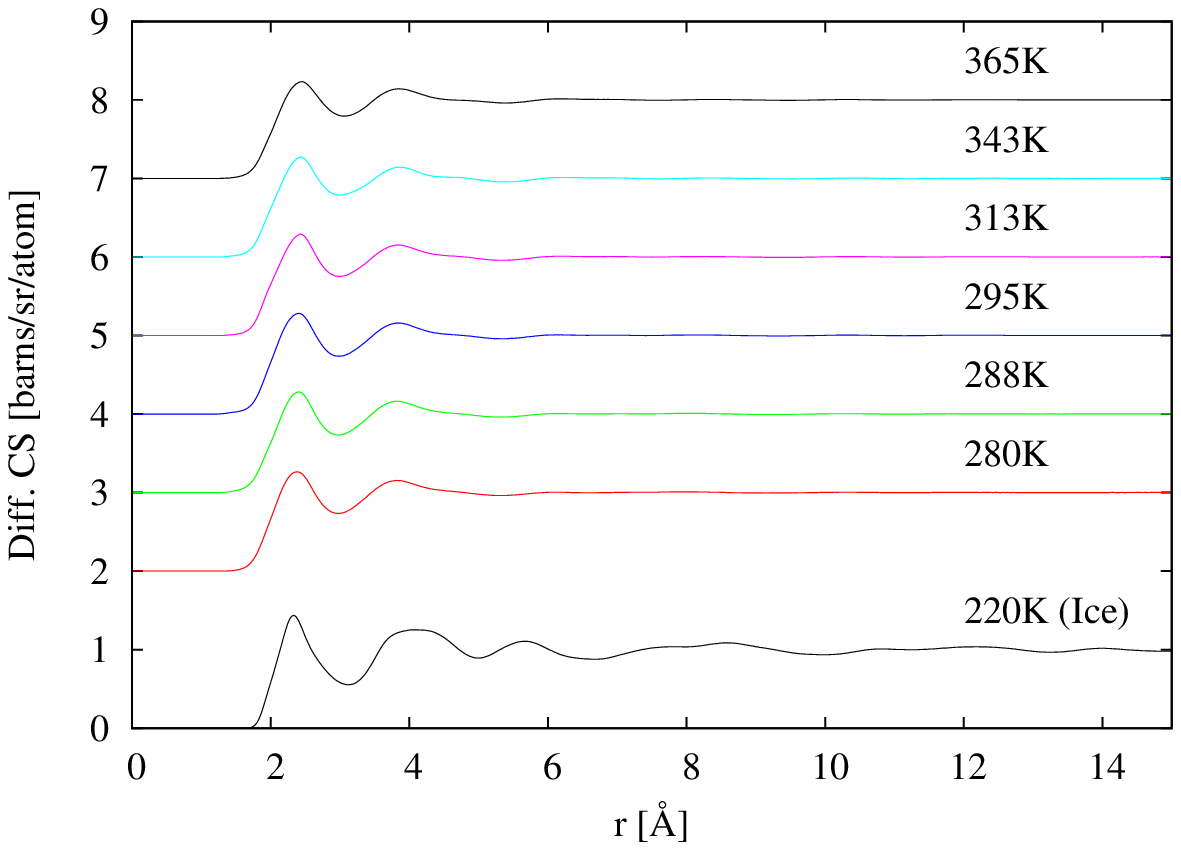} 
\caption{\label{fig:hhgrs}HH radial distribution functions for ice at 220K and water in the temperature range 280-365K, as derived from the EPSR simulations shown in Figs. \ref{fig:h2o22csqfits} and \ref{fig:ice220ksqfits}.}
\end{figure}

The OO, OH and HH radial distribution functions derived from these simulations are shown in Figs. \ref{fig:oogrs}, \ref{fig:ohgrs} and \ref{fig:hhgrs} respectively. The results for the OO distribution for water, Fig. \ref{fig:oogrs}, are closely similar to those originally determined from x-ray scattering by Narten \textit{et al.} \cite{narten1971:1}, namely the first peak moves to slightly larger $r$ with increasing temperature, and the second ``hump'' near $r=4.5$\AA\ becomes increasingly broad and less well defined. But it has not disappeared, even at 365K. This peak is sometimes claimed to indicate the tetrahedral network nature of water. If so then both here and in the earlier work, that network is observed to remain to at least the boiling point of water, 373K. On the other hand the OH and HH functions, Figs. \ref{fig:ohgrs} and \ref{fig:hhgrs} stay remarkably unchanged with increasing temperature. Both functions show the characteristic double peak which is the signature of the strongly directional hydrogen bonding between water molecules, with only very slight broadening and movement with increasing temperature.

For ice at 220K, the peaks in the radial distribution functions are naturally much better defined than in the liquid, particularly at larger $r$ values, where the long range crystalline order becomes apparent. However at low $r$ for all three functions it is striking how broad are all the main peaks: at no point does the density go to zero between the first and second peaks. This points to considerable local \textit{disorder} in the structure of ice, a result which is in agreement with an earlier interpretation of these data \cite{soper2000radial}.

\subsection{\label{derived quantities}Other derived quantities}

As seen in section \ref{experiment} the scattering experiment is related by direct Fourier transform to the site-site radial distribution functions, and these functions are therefore the primary outputs of these experiments. However having set up and performed computer simulations on the same systems, water and ice in the present case, it is natural to probe those simulations for other derived quantities which are not directly accessible through the data. In doing so it is important to realise of course that such derived quantities, because they are not directly constrained by the experiment, may be less reliable than the site-site distribution functions.

Two quantities that are particularly useful for characterising water structure are the atom-atom-atom so-called ``bond angle'' distribution functions and the spatial density functions. The bond angle distribution functions are basically the triple body correlation between three atoms, at least two pairs of which are at or near their near-neighbour distance, forming a local ``bond''. The spatial density function, originally investigated by Kusalik \textit{et al.} \cite{KusalikSvishchev} in computer simulations of water, describes the 3-dimensional correlation function of a pair of water molecules as a function direction as well as distance. Because this is a density as a function of three coordinates, $r$, $\theta$ and $\phi$, it has to be plotted as a surface contour in 2-dimensions. For water and ice it gives a very visual impression of what these materials look like at the molecular level.

\begin{figure}
\includegraphics{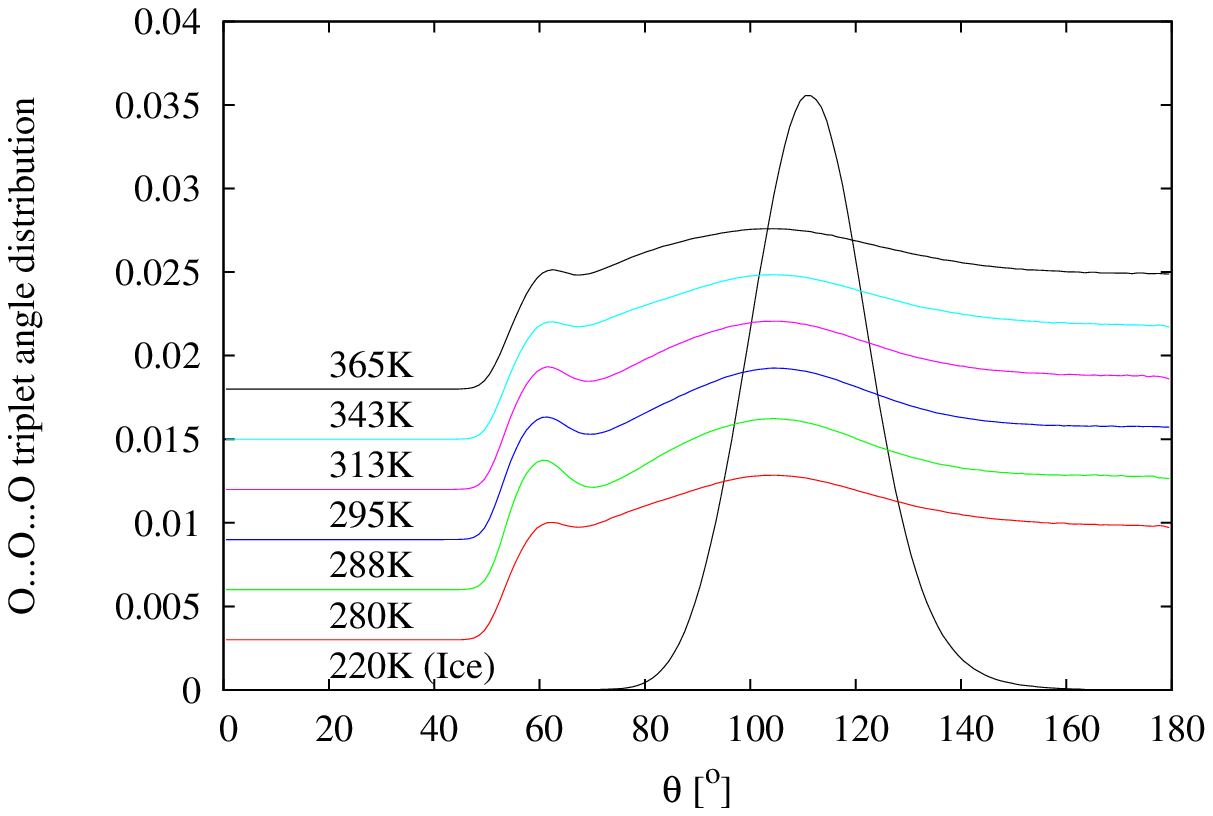} 
\caption{\label{fig:oootriplets}O...O...O triplet included angle distributions for ice at 220K and water in the temperature range 280-365K, as derived from the EPSR simulations shown in Figs. \ref{fig:h2o22csqfits} and \ref{fig:ice220ksqfits}. For convenience the trivial $\sin\theta$ dependence that would occur if the bond angles were completely random is divided out in the representations shown here.}
\end{figure}

For the bond angle distributions, two quantities are particularly helpful for characterising the local order. One of these is the O...O...O included angle distribution where the first and third oxygen atom are near the first neighbour distance from the middle oxygen. (This does not eliminate the case where the first and third oxygen atom are also at the near neighbour distance, creating a near-isosceles triangle with angles near 60$^{\circ}$.) For the present purposes the maximum allowed distance for two O atoms to be classed as bonded was set to 3.15\AA. This is close to the distance at which the O-O coordination number reaches 4 at all the measured temperatures in the liquid. For comparison the same distance is used to calculate this distribution in ice, but note that for the case of ice the O-O coordination number does not reach 4 until the larger distance of $\sim$3.64\AA\ due to the significant width of this peak and the lower density of ice compared to water. In fact using the larger distance did not change the outcome very appreciably. Fig. \ref{fig:oootriplets} shows these estimated distributions for ice at 220K and all six water temperatures. 

We note that, as expected, this triplet angle distribution is strongly peaked near the tetrahedral angle, 109$^{\circ}$, in ice, while in water at all temperatures the same peak is much broader and there are a measureable number of 60$^{\circ}$ triplets in the liquid at all temperatures. It is to be noted that, despite the second peak in the O-O pair distribution function, Fig. \ref{fig:oogrs}, becoming broader with increasing temperature, this triplet angle distribution hardly changes at all with temperatures, implying that the local, tetrahedrally coordinated, water structure is quite resilient to temperature changes in this range. Therefore something other than a simple break up of the tetrahedral network must be happening in liquid water as the temperature rises.

\begin{figure}
\includegraphics{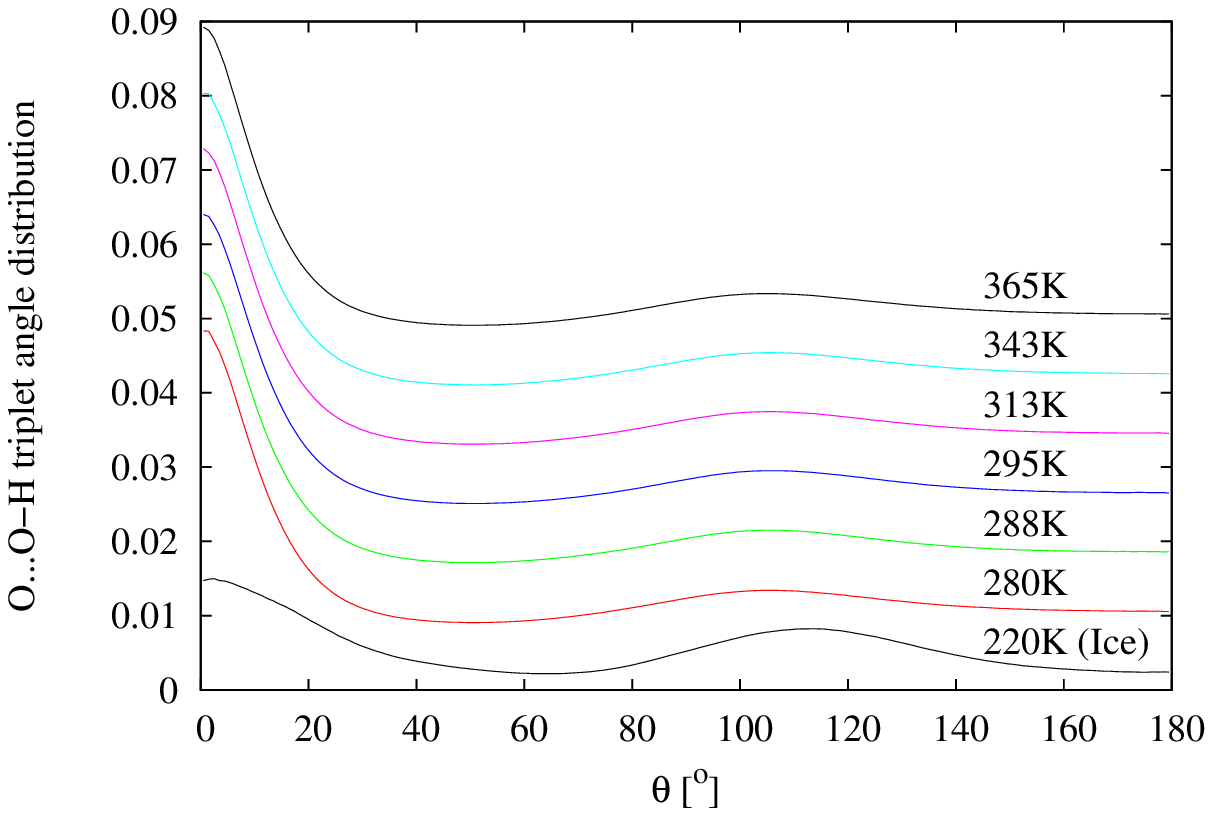} 
\caption{\label{fig:oohtriplets}O...O-H triplet included angle distributions for ice at 220K and water in the temperature range 280-365K, as derived from the EPSR simulations shown in Figs. \ref{fig:h2o22csqfits} and \ref{fig:ice220ksqfits}. For convenience the trivial $\sin\theta$ dependence that would occur if the bond angles were completely random is divided out in the representations shown here.}
\end{figure}

The other bond angle distribution of note is the O...O-H angle distribution which gives an indication of the linearity of the hydrogen bonds in water and ice. For these distributions the O...O distance was set to 3.15\AA\ as above, while the H atom was one of the two hydrogen atoms bonded to each oxygen. A linear hydrogen bond would give a sharp peak in this function at $\theta=0^{\circ}$ and second peak at $\sim 105^{\circ}$ corresponding to the second hydrogen on a water molecule bonding to a third water molecule if the first is pointing directly at a neighbouring oxygen atom. Fig. \ref{fig:oohtriplets} shows these hydrogen bond angle distributions for ice and water as found in the present simulations.

Here it can be seen that the first peak in these distributions is much sharper, and therefore more linear, in the liquid than in ice, implying the hydrogen bond in the liquid is stronger and better defined than in the solid. On the other hand, the second peak near 110$^{\circ}$ is much weaker in the liquid than in the solid. This has to occur because in ice each hydrogen atom has a choice of four neighbouring oxygen atoms to bond to with equal probability, leading to the well known proton disorder in ice Ih. On the other hand in liquid water, the presence of non-tetrahedral water molecules in the first coordination shell means this second hydrogen may lack a clear bonding site if the first is bonded, which frees up the first hydrogen to form a stronger bond with its neighbour. This fundamental asymmetry in water structure is almost certainly an important clue to the anomalous thermodynamic behaviour of water as a function of temperature and pressure.

\begin{figure}
\includegraphics[scale=0.13]{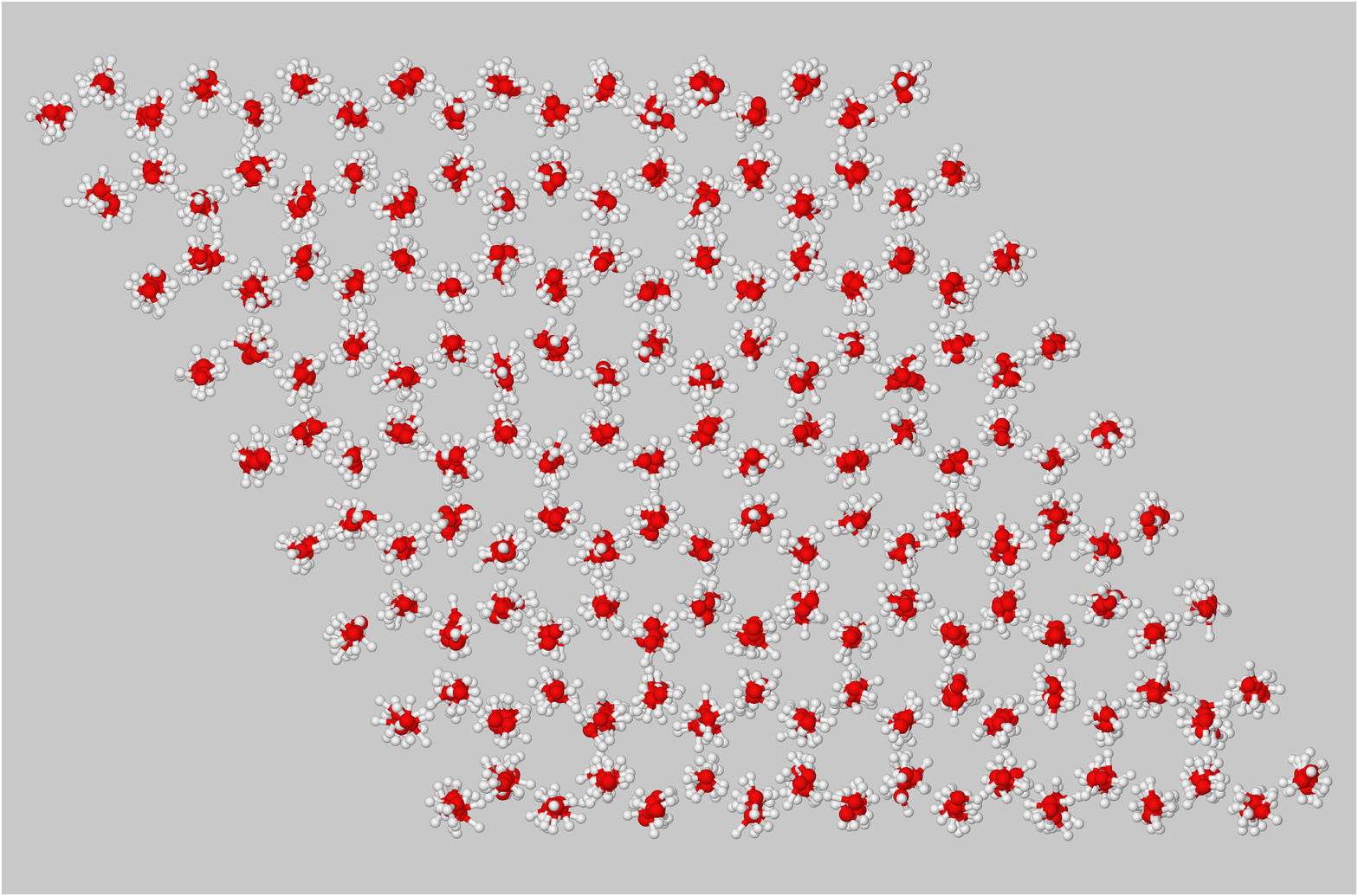} 
\caption{\label{fig:icesimbox}Ice Ih EPSR simulation box, viewed along the crystallographic $c$-axis. The significant proton disorder is apparent in this graphic, alongside the long range order of the oxygen lattice. }
\end{figure}

The same message can be captured by plotting the simulation box for the crystalline ice structure refinement, Fig. \ref{fig:icesimbox}, here viewed along the crystallographic $c$-axis. The oxygen atoms are mostly near their crystallographic positions, but the hydrogen atoms are clearly significantly disordered. Attempts to make the hydrogen bonds more linear as might be expected in a simple view of ice Ih structure failed: whenever this was done, the fit to the scattering data became markedly worse.

The spatial density function for water is calculated by placing each water molecule in the simulation box at the centre of the local coordinate system, with the $z$-axis running through the oxygen atom and bisecting the two hydrogen atoms, and the $z$-$y$ plane being defined by the plane of the water molecule. The positions of neighbouring water molecules are then recorded as a function $r,\theta,\phi$ with respect to this coordinate system. The distribution is averaged over all the water molecules in the box and over a number of configurations of the simulation box - typically around 100. In order to plot the surface contour one needs to choose the distance range over which the distribution is to be plotted, and also what fraction of all the molecules in that distance range will be included inside the surface. In this way plots where the average density of molecules is different due to changes of temperature or pressure can be compared in a meaningful way. For the present examples the distance range was set from 0 - 5\AA\ and the fraction of all molecules inside the contour surface was set to 25\%. A detailed explanation of how these plots are generated can be found at \cite{epsrmanual}.

\pagebreak
\begin{figure}
\begin{tabular}{c}
(a) Water at 365K \\
\includegraphics[width=0.4\textwidth]{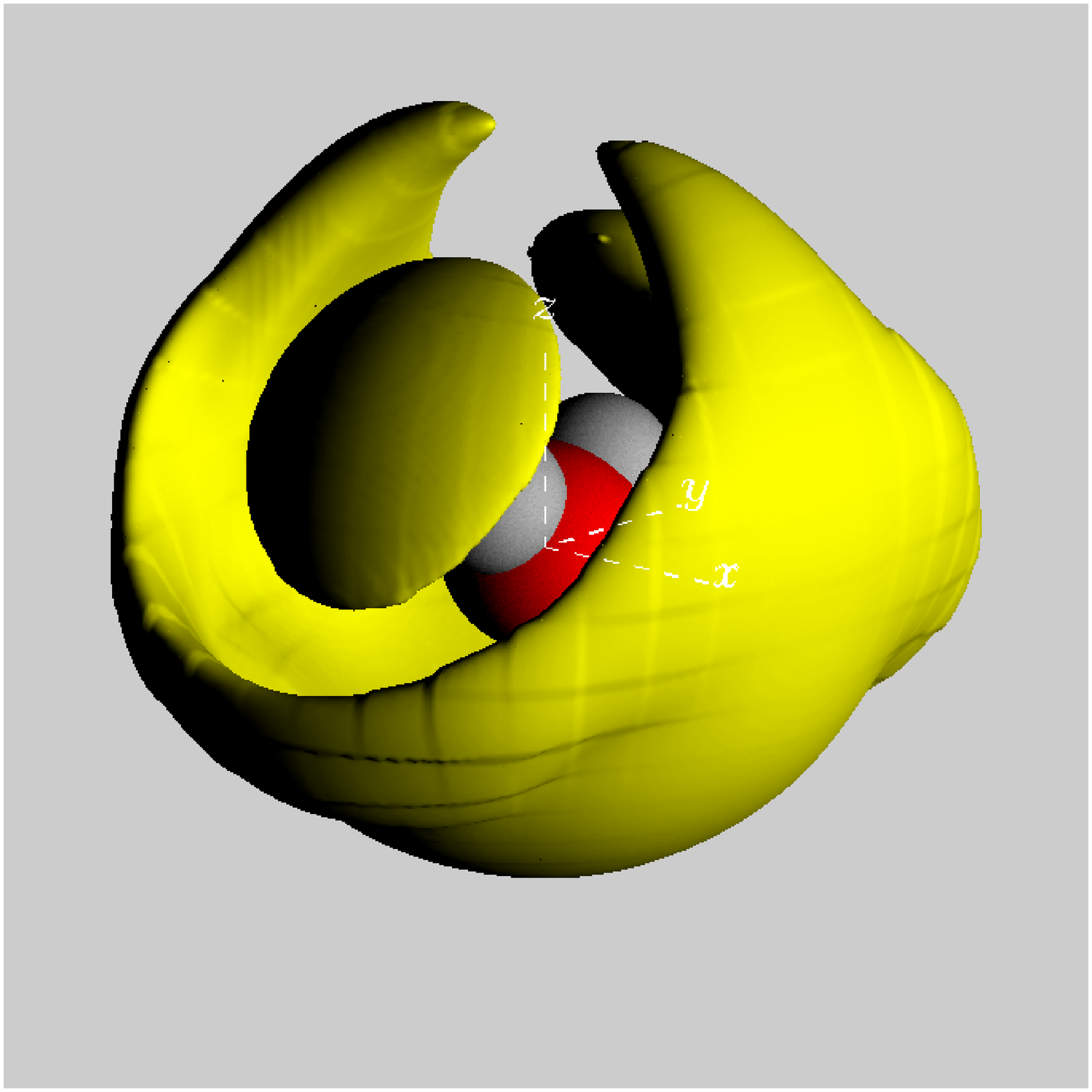} \\
(b) Water at 280K \\
\includegraphics[width=0.4\textwidth]{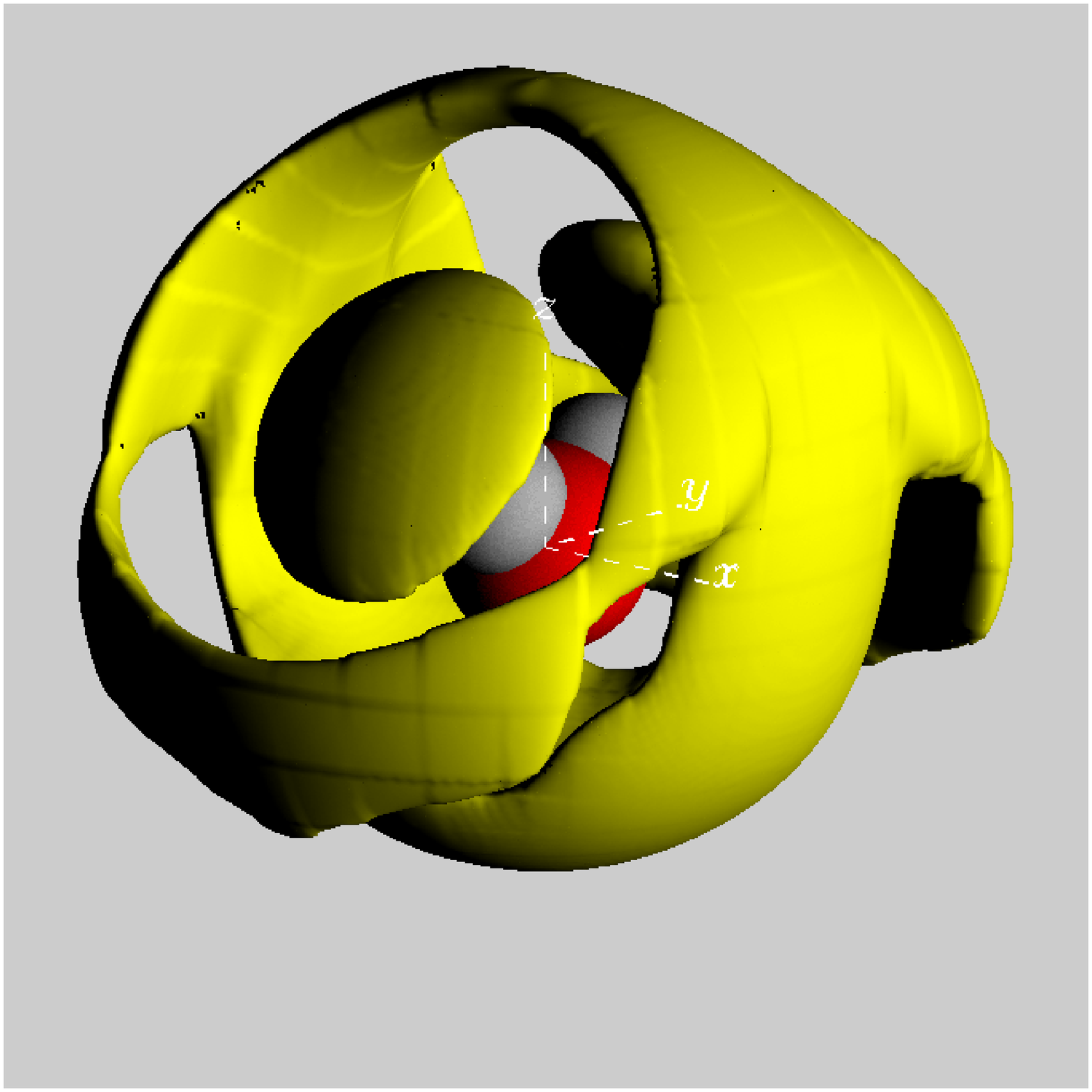} \\
(c) Ice Ih at 220K \\
\includegraphics[width=0.4\textwidth]{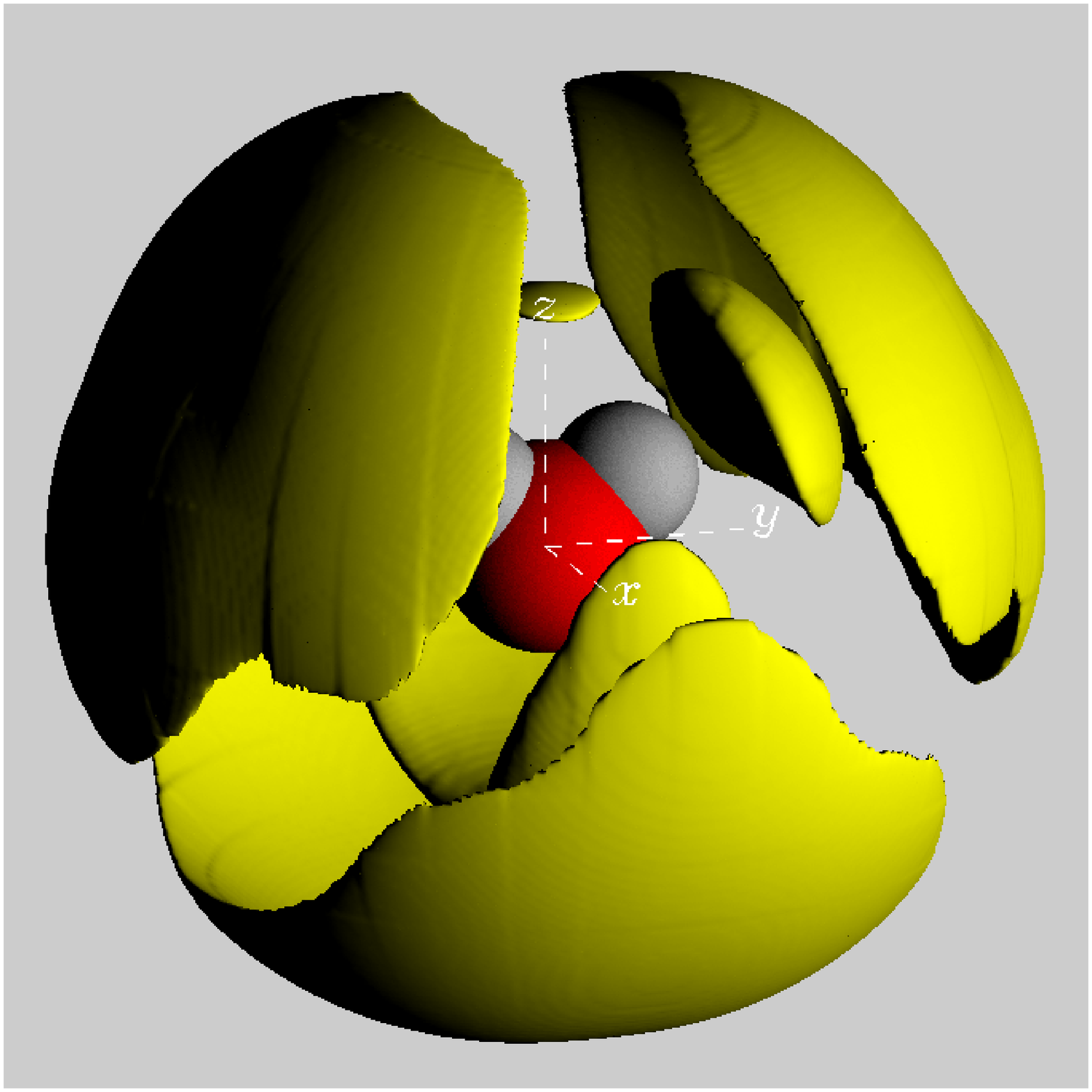} \\
\end{tabular}
\caption{Spatial density functions for water at 365K (a), water at 280K (b), and ice Ih at 220k(c). The surface contours enclose the top (most dense) 25\%\ of water molecules within 5\AA\ from a central water molecule. The faint white dashed lines show the coordinate axes of the central water molecule.}
\label{fig:sdfs} 
\end{figure}

Fig. \ref{fig:sdfs} (a) - (c) show the calculated spatial density functions for water at 365K (a), water at 280K (b) and ice 1h (c). The very marked difference in local order between liquid water, (a) and (b), and crystalline ice, (c), is apparent in these plots. In both cases the inner core of water neighbours appears to be very similar between water and ice, with a pronounced lobe of intensity opposite each hydrogen atom of the central molecule, and a broader, more continuous band of intensity underneath. The second shells however differ completely between the solid and liquid forms. In the crystalline form the second shell is essentially a mimic of the first shell, about twice as far out, but in the liquid the second shell merges with the first shell in some directions, but at lower temperatures (280K) is in almost exact antiphase with the first shell. The second shell then partly collapses into the first shell as the temperature is raised to 365K, but note that the local order in the first shell barely changes in this temperature range. Note that this collapse of the second shell occurs despite the decrease in density as the temperature is raised, Table \ref{tab:simboxes}. A similar pattern is observed when low temperature water is pressurised and when the density certainly does increase \cite{soper2000structures}. Hence the local order in the first shell around a water molecule in ice at 220K and in water near the boiling point remains virtually unchanged: all the changes occur in the second shell. In particular when going from ice to water, then subsequently heating the water, this second shell collapses from being a replica of the first shell in ice Ih to being in antiphase with the first shell in the liquid and at higher temperatures.

\section{\label{discussion}Discussion}

The results of the EPSR analysis shown in this paper reveal aspects of the structure of water and ice that are not often alluded to. In particular we find that the \textit{local} (first neighbour) order in water is mostly unaffected on heating the liquid from 280K to 365K, Figs. \ref{fig:oootriplets},\ref{fig:oohtriplets} and \ref{fig:sdfs} (a) and (b). Instead it is the second, ``interstitial'' shell that readily collapses inwards on heating, implying increased bending of the hydrogen bonds between triplets of water molecules, although the effect is subtle, Fig. \ref{fig:oootriplets}.

In ice, although disordered to some extent, the water molecule oxygen atoms lie on a well-defined tetrahedral lattice, Fig. \ref{fig:icesimbox} with no sign of the interstitial water molecules found in the liquid. On the other hand the water hydrogen atoms in ice are surprisingly disordered. In particular the O...O-H triplet angle distribution is significantly broader in the region 0 - 60$^{\circ}$ than in water. On the other hand in the region of 60-180$^{\circ}$ the same distribution in ice is much sharper than in water. It can be surmised that these differences arise from the different environments that a water molecule finds itself in in the two materials. In ice the neighbouring four water molecules are arranged rather regularly on a tetrahedral lattice: the hydrogen atoms of the central molecule are therefore as equally likely to bond to any one of these four molecules. They therefore find themselves in a state of equally ``divided loyalities'' with the neighbouring water molecules, making it difficult to form linear, and therefore presumably stronger, hydrogen bonds with any one neighbour. In the liquid however, if a linear bond does form, then it is likely there will less competition from the neighbouring water molecules, due to their more random arrangements, so helping to preserve the linearity. This idea would explain why the second peak in the O...O-H distribution is sharper in ice compared to water: in ice any individual bond is likely to be less linear than in the liquid, but overall the bonding environment is more favourable, leading to a significantly larger hydrogen bonding energy for the structure as a whole.

This discussion must also take account of the significant zero point energy effects that affect hydrogen bonding in water. The computer simulations given in \cite{fanourgakis2009fast} show clearly how significant quantum zero point disorder affects both the structure and dynamics of water. This zero point disorder derives primarily from the way the proton is bound to the oxygen atom in a water molecule, influenced of course to some extent by the inter-molecular bonding. Zero point disorder is by its very nature nearly temperature independent, and so is likely to significantly affect proton disorder in crystalline ice as well. 

Given that the distributions derived in this paper are arrived at by a process of measuring data from crystalline ice on a low-resolution liquids diffractometer, then subjecting them to a fairly extensive, but far from perfect, computer simulation regime, it is tempting to propose that the results shown here are simply artefacts of the way the data have been handled. However our view is that this attitude is too simplistic. The widths of the peaks, and hence the strength of correlations, in real space are determined by the decay of intensity with increasing $Q$ in reciprocal space - the Debye-Waller factor. The low resolution of the diffractometer means that Bragg peaks in reciprocal space cannot be individually identified in the present data at large $Q$, so that only the envelope of those peaks can be determined. If significant broadening effects were imposed on the data due to the resolution of the diffractometer they would affect the liquid water results as much as they affect the ice results. Moreover this resolution function $is$ built into the EPSR simulations described here, which should at least partly ameliorate for the loss of resolution in the real space functions that would otherwise occur from lack of $Q$-space resolution. Although detailed reproduction of all the Bragg intensities here is inexact, the gualitative trends in these intensities with increasing $Q$ are reproduced correctly by the current EPSR simulations. Moreover a recent computer simulation of the transformation from amorphous to crystalline ice \cite{Limmer23052014} suggests a degree of disorder in ice Ih not incompatible with the present estimate. There have also been extensive discussions on the degree of disorder in ice based on crystallographic information - see for example \cite{kuhs1986water, kuhs1986oxygen} - and the present results are not  inconsistent with those conclusions, although, because radial distribution functions for ice are rarely calculated or shown, it is difficult to make direct comparisons.

Another issue that is often raised when discussing ice is the Bernal-Fowler ice rules \cite{bf}, which state that there should be one hydrogen atom between each pair of oxygen atoms in the lattice, either covalently bonded or hydrogen bonded to one of the oxygen atoms, at any one time, and that each oxygen must have two covalently bonded hydrogen atoms. This idealisation of the real lattice will be difficult to adhere to strictly since the bond angle between the two covalently bonded hydrogen atoms in ice is believed to be around 107$^{\circ}$ \cite{kuhs1987geometry}, that is, less than the 109.47$^{\circ}$ required for precise tetrahedral bonding. Quantum zero point effects as discussed above will make this strict adherence even less likely in an instantaneous sense, though on averaging over the positions of the hydrogen atoms, the ice rules should still apply. The question then is: does the current EPSR model of ice Ih also adhere to the ice rules? Clearly the condition that each water molecule have two covalently bonded hydrogen atoms is satisfied, since that is the way the water molecules have been defined within the EPSR framework. On the other hand the condition that there is only one hydrogen atom between each pair of oxygen atoms will be more difficult to enforce due to the significant hydrogen atom disorder already discussed. To ensure the likelihood of two hydrogen atoms occurring in the same O...O bond was kept acceptably low, a minimum H...H intermolecular distance of 1.7\AA\ has been imposed in the current EPSR simulations, Table \ref{tab:minimumdistances}. In any case such contacts would occur only for short periods in practice due to the Coulomb repulsion between the hydrogen atoms in this model.

\section{\label{conclusion}Conclusion}

The foregoing has described some empirical potential structure refinement simulations of liquid water in the temperature range 280K to 365K, that is from near the freezing point to near the boiling point, and also of crystalline ice at 220K. The local, first shell structure in this temperature range is found to be remarkably resilient to the change of temperature, with a tetrahedral-like first shell, near linear O...O-H hydrogen bonds, a O...O...O triplet angle distribution which changes only slowly with increasing temperature. On the other hand the second neighbour, interstitial, shell is quite sensitive to changes in temperature, collapsing inwards as the temperature increases, presumably due to subtle changes in the O...O...O and O...O-H triplet distributions. These ideas are borne out in the EPSR ice simulations where the complete absence of an interstitial shell is marked by O...O...O triplet angles centred entirely on the tetrahedral angle at 109.47$^{\circ}$. Interestingly the O...O-H distribution is markedly less linear in the crystal compared the liquid, probably because of the much more disordered arrangement of neighbouring water molecules in the liquid. The present results for the radial distribution functions of ice Ih differ somewhat from those shown in a previous publication \cite{soper2000} because in the present case the water molecules have been deliberately constrained to lie near their crystallographic positions, while previously they were free to roam throughout the simulation box. In that case the simulation model would probably demonstrate excessive diffuse scattering compared to what is measured \cite{kuhs1986water} due to the substantial oxygen disorder. An interesting exercise here would be to calculate the diffuse scattering from the present EPSR model and see how it compares with the same experiments.

\section{Acknowledgments}
The author is indebted to Sam Callear for assistance with the x-ray scattering experiments, to Daniel Bowron for assistance with the neutron scattering experiments, and to Werner Kuhs for information about earlier ice crystallography.

%\begin{figure}
%\includegraphics{foo}     % includes figure foo.eps
%\caption{Description of the figure.}
%\end{figure}

%\section{Examples}

%\subsection{Tables}
%Tables~\ref{tab:pricesI}, \ref{tab:pricesII} and~\ref{tab:pricesIII}
%inserted at this point.

%\begin{table}
%  \caption{Prices of important items.}
% \label{tab:pricesI}
  %\begin{tabular}{rcl}
    %\hline
      %Ice-cream      & 1500  & lire    \\
      %More ice-cream & 15000 & lire    \\
      %Crocodile      & 1500  & dollars \\
    %\hline
      %Phone call     & .25   & dollars \\
      %X-Men          & 1.25  & dollars \\
      %Dollar         & 1     & dollars \\
    %\hline
  %\end{tabular}
%\end{table}

%\subsection{Mathematics}
%Here is a lettered array~(\ref{e.all}), with eqs.~(\ref{e.house})
%and~(\ref{e.phi}):
%\begin{eqnletter}
 %\label{e.all}
 %\drm x_\sy{F} & = & 1.2\cdot10^3\un{cm}, \qquad
 %                    \tx{where\ } \sy{F} = \tx{Fermi}    \label{e.house}\\
 %\phi_i        & = & i\pi                                \label{e.phi}
%\end{eqnletter}

%\subsection{Citations}
%We're almost done, just some citations~\cite{ref:apo}
%and we will be over~\cite{ref:pul,ref:bra}.

%\begin{thebibliography}{0}
%\bibitem{ref:apo} \NAME{Einstein A. \atque Fermi E.}
  %\IN{Phys. Rev. A}{13}{1999}{12};
  %\SAME{69}{999}{1666}.
%\bibitem{ref:pul} \NAME{Newton I.}
  %preprint INFN 8181.
%\bibitem{ref:bra} \NAME{Bragg~B.}
  %\TITLE{Complete Works}, in \TITLE{Workers Playtime}, edited by \NAME{Tizio A. \atque Caio B.} (Unexeditor, %Bologna) 1997, pp.~1-10.

%\bibliographystyle{varenna}
%\bibliography{VarennaAKS,FrontiersArticle,Frontiers,NewWaterStructureII}

\end{document}